%% file: FreeEnergy.tex
\documentclass[12pt]{article}   
\usepackage{geometry,bbm}   
\usepackage[toc,page]{appendix}             		
\usepackage{graphicx}
\usepackage[usenames,dvipsnames]{color}
\usepackage{slashed}					
\usepackage{amssymb,amsmath}
\usepackage{hyperref}
\usepackage{mathtools}
\usepackage{multirow}
\usepackage[table]{xcolor}

\def\){\right)}
\def\({\left( }
\def\]{\right] }
\def\[{\left[ }

\newcommand{\tx}{\text}

\newcommand{\w}{\wedge}

\newcommand{\ea}{\end{align*}}
\newcommand{\ba}{\begin{align*}}
\newcommand{\bi}{\begin{itemize}}
\newcommand{\ei}{\end{itemize}}

\newcommand{\ben}{\begin{enumerate}}
\newcommand{\een}{\end{enumerate}}

\newcommand{\vol}{\text{vol}}

\include{macros}
\numberwithin{equation}{section}
\begin{document}

\begin{titlepage}

\begin{flushright}
SISSA  53/2016/FISI
\end{flushright}
\bigskip
\def\thefootnote{\fnsymbol{footnote}}
\begin{center}

{\LARGE
{\bf
A note on sphere free energy of \\ \vskip 20pt $p$-form gauge theory and Hodge duality}}
\end{center}

\bigskip
\begin{center}
{\large
Himanshu Raj}
\end{center}

\renewcommand{\thefootnote}{\arabic{footnote}}

\begin{center}
\vspace{0.2cm}
{SISSA and INFN - 
Via Bonomea 265; I 34136 Trieste, Italy\\}

\vskip 5pt
{\texttt{hraj@sissa.it}}

\end{center}

\noindent
\begin{center} {\bf Abstract} \end{center}
\noindent
We consider a free $p$-form gauge theory on a $d$-dimensional sphere of radius $R$ and calculate its free energy. We perform the calculation for generic values of $p$ and obtain the free energy as a function of $d, p$ and $R$. The result contains a $\log R$ term  with coefficient proportional to $\(2p+2-d\)$, which is consistent with lack of conformal invariance for $p$ form theories in dimensions other than $2p+2$. We also compare the result for $p$-form and $(d-p-2)$-form theory which are classically Hodge dual to each other in $d$ dimensions and find that they agree for odd values of $d$. Instead, for even $d$, we find that the results disagree by an amount that is consistent with the reported values in the literature.

\vspace{1.6 cm}
\vfill

\end{titlepage}

\newpage


\section{Introduction}
\label{section1}
Sphere free energy plays an important role in the study of Renormalization Group (RG) in $d$-dimensional Quantum Field Theory (QFT). If an RG connects a unitary UV Conformal Field Theory (CFT) to a unitary IR CFT, then a certain positive quantity defined on the space of CFTs, decreases monotonically. This is the well known $c$- , $F$- and $a$-theorem in $d=2,3,4$ respectively \cite{Zamolodchikov:1986gt,Komargodski:2011vj,Komargodski:2011xv,Jafferis:2011zi,Casini:2012ei}.  Irrespective of the dimension $d$, this quantity can be extracted from the free energy of the theory defined on a sphere. Motivated by the similarity of such monotonicity theorems in even and odd dimensions (both of them can be formulated in terms of sphere free energy) it was conjectured in \cite{Giombi:2014xxa} that these theorems are special cases of a more general one, valid in continuous range of dimension for which
\be\label{genFthem}
\wt F_{\tx{UV}}(d)>\wt F_{\tx{IR}}(d)~,
\ee
where $\wt F(d)=-\sin(\p d/2) F(d)$ and $F(d)$ is the sphere free energy which is defined as $F(d)= -\log Z(S^d)$. The quantity $\wt F$ which turns out to be a smooth function of $d$, interpolates between the $a$-anomaly coefficient in even dimensions and the sphere free energy in odd dimensions. Thus the inequality \eqref{genFthem} (dubbed as the ``Generalized $F$-theorem" in \cite{Fei:2015oha}) smoothly interpolates between the corresponding inequalities in even and odd dimensions. 

The value of $\wt F$ has been calculated for several theories. Some interesting examples include a free conformal scalar, spin $1/2$ fermions, $U(1)$ Maxwell theory, the interacting $O(N)$ model, double trace perturbations of large-N CFTs and conformal QED \cite{Giombi:2014xxa,Fei:2015oha,Giombi:2015haa,Bashmakov:2016pcg,Tarnopolsky:2016vvd}.

In this note we calculate $\wt F$ for a free massless $p$-form gauge theory on a $d$-dimensional sphere of radius $R$. Our calculation is based on the method prescribed in section 2 of \cite{Giombi:2015haa} where $\wt F$ was calculated for $p=0\tx{ and }1$. We extend this calculation to generic $p$ and present explicit results for $p=0,1,2,3,4$. As a check of these results we calculate the $a$-anomaly coefficient of these theories in $d=2p+2$ where they are conformal and we find agreement with known values in the literature. Then we compute $\wt F$ for $p$-form gauge theories in odd dimensions. In particular, we present the result for a 2-form on $S^5$ and find that it matches exactly with the corresponding value for a 1-form on $S^5$, obtained in \cite{Giombi:2015haa} ,which is expected to hold via Hodge duality. We also check dualities for other values of $p$. For odd $d$ we find perfect agreement with classical expectations. On the contrary, for even $d$, we find that there is mild disagreement in $\wt F$ for the classically Hodge dual pair of theories. This discrepancy has been noticed in the literature. Particularly, in 4 dimensions it is known that the contribution of a scalar field and a 2-form to the trace of the energy-momentum tensor differ \cite{Bern:2016,Duff:1980qv,Larsen:2015aia}. The results from our general formula for $\wt F(d)$ in \eqref{freeen2}, which is valid in continuous dimension, are in complete agreement with the reported values in the literature. 


\section{Free energy of a $p$-form gauge theory on $S^d$}
\label{section2}
The action for a free massless Abelian $p$-form gauge field $A_p$ on a curved manifold is
\be\label{action}
S=\frac{1}{2} \int_{\cm_d}  ~ F_{p+1} \w \star F_{p+1}~,~~~ F_{p+1}=dA_p~.
\ee

This is a scale invariant theory as there are no dimensionful parameters in the action \eqref{action}. However, it is conformally invariant only in $d=2p+2$ dimensions\footnote{One way to see this is to calculate the trace of the energy-momentum tensor and observe that it vanishes in $d=2p+2.$ However, this argument is not sufficient since one may possibly find `improvement' terms for the energy-momentum tensor. One has to further show that the energy-momentum tensor cannot be improved. See ref. \cite{ElShowk:2011gz} for an enlightening discussion for $p=1$.}.

This theory has a large gauge redundancy that needs to be fixed before performing the path integral. The number of degrees of freedom needed to describe the theory in a Lorentz-invariant manner is $d \choose p$. However, the number of  propagating degrees of freedom are $d-2 \choose p$ (which implies that for $p>d-2$ there are no propagating degrees of freedom). Hence, for a proper path integral treatment, we should impose ${d \choose p} -{d-2 \choose p}$ gauge-fixing conditions. This problem has been well studied in the literature (see \cite{Cappelli:2000fe} and references therein).

The action \eqref{action} is invariant under the gauge transformation $A_p\to A_p+dA_{p-1},$ where $A_{p-1}$ is a form of degree $(p-1)$. Therefore, one has to introduce in the action gauge-fixing terms for $A_p$ and ghost fields which are also form-fields but of degree less than $p$. The action for the ghost field has its own gauge invariance and therefore one has to introduce further gauge-fixing terms and ghost fields for the ghost action and so on until one arrives at the action for the scalar ghost. The resulting gauge-fixed action contains a tower of ghosts with alternating statistics.

On a $d$-dimensional sphere, the Euclidean path integral over the gauge-fixed action yields the following partition function for the $p$-form gauge theory \cite{Cappelli:2000fe}
\begin{align}\label{partitionfunc}
Z_p= \[\frac{1}{\det_T \D_p}\frac{\det_T \D_{p-1}}{\det_T \D_{p-2}} \dots \(\frac{\det_T \D_{1}}{\det' \D_{0}}~\vol(S^d)\)^{(-1)^p}~\]^{1/2}~,
\end{align}
where the measure in the path integral is defined such that there are no factors of $2\p$ in evaluation of the Gaussian functional integral. $\D_p$ is the Hodge-de Rham operator acting on a form of degree $p$. The subscript $T$ means that the determinant is taken over the space of coexact (or transverse) $p$-forms. The prime on scalar determinant means that the scalar zero mode should be removed in the evaluation of the determinant. The volume factor,
\be
\vol(S^d)= \frac{2\p^{\frac{d+1}{2}}}{\Gamma(\frac{d+1}{2})}R^d\equiv \Omega_d R^d~,
\ee
is a consequence of the regularization of the zero mode of the scalar ghost and the lack of zero modes for $p\ge 1$ forms on a sphere\footnote{The scalar action $\int (\pa \f)^2$ has a shift symmetry by a constant $\f \to \f +\f_0$ which should be fixed by a Fadeev-Popov argument. The volume factor appears upon inserting the delta function in a dimensionless manner. We thank A. Cappelli for this explanation.}.

The determinants in \eqref{partitionfunc} can be evaluated in the basis of transverse $p$-form spherical harmonics. The eigenvalues and the corresponding degeneracies of these harmonics are given by \cite{Cappelli:2000fe, Elizalde:1996nb}
\begin{align}\label{eigenval}
\l_{p,l} &=\frac{1}{R^2}(l+p)(l+d-p-1)~,\NO \\
g_{p,l} &=\frac{(2l+d-1)\G(l+d)}{\G(p+1)\G(d-p)\G(l)(l+p)(l+d-p-1)}~.
\end{align}
In these formulas $l\geq 0$ for $p=0$ and $l>1$ otherwise. However since the scalar zero mode is not included in \eqref{partitionfunc} we will henceforth take $l>0$ for all $p$. In the remainder of this section we will evaluate the ratio of the determinants in \eqref{partitionfunc} on $S^d$ in dimensional regularization. 

To proceed with the calculation we find it convenient to organize the terms in the sphere free energy, as follows
\begin{align}\label{freeen1}
F_p(d) &=-\log Z_p(S^d)=\sum_{n=0}^{p}(-1)^{p-n} {p+1\choose p-n}f_n(d)-\frac{(-1)^{p}}{2}\log \(\vol(S^d)\)~,
\end{align}
where
\begin{align}\label{pformcont}
f_p &(d)=\frac12\sum_{n=0}^p {p \choose n}~{\log \det}_T \D_n~.
\end{align}
The logarithm of the determinant can be written as
\begin{align}\label{logdet}
I_p\equiv \log \tx{det}_{T}\D_p&=\sum_{l=1}^\infty g_{p,l}\log \l_{p,l}~\NO\\
&=\sum_{l=1}^\infty g_{p,l}\log \[(l+p)(l+d-p-1)\]+(-1)^{p}\log \(\m_0R\)^2~,
\end{align}
where we have inserted an arbitrary scale $\m_0$ in the logarithm to soak up the dependence on the radius of the sphere.  In the evaluation of the last term in \eqref{logdet} we have used the following result for the sum over the degeneracies $g_{p,l}$,
\be\label{sumg}
\sum_{l=1}^\infty g_{p,l}=-\cos (p \p)=(-1)^{p+1}~.
\ee
The above sum converges for sufficiently negative $d$ which can then be analytically continued to positive $d$. Before proceeding let us first extract the radius dependence of the free energy. We write \eqref{freeen1} as
\begin{align}
F_p(d)=F^0_p(d) + F^{(R)}_p(d)~,
\end{align}
where $F^0_p(d)$ is a radius independent function of $d$. Using the dimensionally regularized sum in eq. \eqref{sumg} we find that 
\be
F^{(R)}_p(d)=(-1)^p\(p+1-\frac d2\)\log \(R\m_0\)~.
\ee
For $p=0$ or $1$ and $d=3$ this logarithmic term has coefficient $-\frac 12$ which was previously obtained in \cite{Giombi:2015haa, DiPietro:2014bca}. In odd dimensions the coefficient of the log is invariant under duality. From the above formula we see that in $d=2p+2$ dimensions the dependence of the free energy on the scale $\m_0$ vanishes. This is consistent with the fact that such a theory is conformal as previously remarked.

To proceed further with the evaluation of $F_p(d)$ we make use of the identity
\be
\log y=\int_0^\infty \frac{dt}{t} \(e^{-t}-e^{-yt}\)~.
\ee
Using \eqref{sumg} the radius independent part of the sum $I_p$, denoted below as $I_p^0$, can be written as
\begin{align}\label{logdet1}
I_p^0 =-\int_0^\infty \frac{dt}{t} \(\(-1\)^{p}2e^{-t}+\[\sum_{l=1}^\infty g_{p,l}\(e^{-(l+p)t}+e^{-(l+d-p-1)t}\)\]\)~.
\end{align}
The infinite sum over the eigenvalues in the square brackets can be evaluated as a finite sum in terms of hyperbolic functions of $t$. After some algebra we find the following result
\begin{align}
\sum_{l=1}^\infty g_{p,l}\(e^{-(l+p)t}+e^{-(l+d-p-1)t}\)=&4\m(p)\cosh\(\frac t2 (d-2p-1)\)\frac{e^{-\frac d2 t}}{\(1-e^{-t}\)^d}\NO\\
&+(-1)^{p+1}\(1+e^{-t (d-2p-1)}\)~,
\end{align}
where the finite sum $\m(p)$ is defined as 
\be
\m(p)=\sum_{n=0}^{p}(-1)^n\frac{\G(d+1)}{\G(d-p+n+1)\G(p-n+1)}\cosh\(\frac t2 (2n+1)\)~.
\ee
To perform an analytic evaluation of the $t$ integral in \eqref{logdet1} it is convenient to use the identity
\be
\frac 1t=\frac{1}{1-e^{-t}}\int_0^1 du~ e^{-ut}~.
\ee
After some straightforward algebra we obtain the following general result for the radius independent part of $f_p(d)$, denoted below as $f^0_{p}(d)$. For $p>0$ we have
\begin{align}\label{pformcont1}
f_p^0(d)&=\int_0^1 du\[\frac{\G(1+u)\G(d-u-2p)}{2p!\G(d)\sin\(\frac{\p d}{2}\)}\sin\(\frac12 \p\(d-2u\)\)q_p(u,d)+G_p(u,d)\]~,
\end{align}
while for $p=0$ the result is
\begin{align}\label{scalarf0}
f_0^0(d)&=\int_0^1 du\[\frac{(2u-d)\G(d-u)\G(u)\sin\(\frac12 \p\(d-2u\)\)}{2\G(d+1)\sin\(\frac{\p d}{2}\)} +\frac{1}{2u} \]-\frac12\log(d-1)~.
\end{align}
The function $G_p(u,d)$ in \eqref{pformcont1} is defined recursively through the following relation:
\be
G_p(u,d)=(-1)^{p+1}\frac{2p-d}{(2p-d)^2-u^2}+\sum_{k=1}^{p-1}(-1)^{k+1}{p-1\choose p-k-1}G_{p-k}(u,d)~,
\ee
and the function $q_p(u,d)$ is a polynomial of degree $2p$ in both $u$ and $d$. Appendix \ref{app1} contains the expression for $q_p(u,d)$ for $p=1,2,3,4$. We have not been able to obtain a compact expression for $q_p(u,d)$ for arbitrary $p$. Plugging this result in the expression for the free energy \eqref{freeen1} we arrive at the main result of our paper,
\begin{align}\label{freeen2}
F_p(d) =\sum_{n=0}^{p}&(-1)^{p-n} {p+1\choose p-n}f_n^0(d)+\frac{(-1)^{p+1}}{2}\log \(\Omega_d\)\NO\\
&+(-1)^p\(p+1-\frac d2\)\log \(\m_0 R\)~.
\end{align}
From \eqref{pformcont1} and \eqref{scalarf0}, we see that the above expression has simple poles at even integer values of $d$. However, the quantity $\wt F_p(d)=-\sin(\p d/2) F_p(d)$ is finite for even $d$ and is a smooth function in continuous range of dimension. For CFTs, the quantity $\wt F_p(d)$ smoothly interpolates between $(-1)^{d/2}(\p/2)$ times the $a$-anomaly coefficient in even dimensions and $(-1)^{(d+1)/2}$ times the sphere free energy in odd dimensions \cite{Giombi:2014xxa}.

As a check of these results let us calculate below the value of $\wt F$ for various values of $p$. In $d=2p+2$ dimensions, where the $p$-form theory is conformal, we have
\begin{align}\label{f1}
\wt F_1(4) &=-\frac{\p}{12} \int_0^1 du~   (u-1) (12 - 11 u - u^2 + u^3)=\frac{\p}{2} \cdot \frac{31}{45}~,\\[5pt]
\label{f2}
\wt F_2(6) &=\frac{\p}{240} \int_0^1 du~    (u-1) (-360 + 314 u + 41 u^2 - 39 u^3 - 4 u^4 + 2 u^5)\NO\\[5pt]
&=\frac{\p}{2} \cdot \frac{221}{210}~,\\[10pt]\label{f3}
\wt F_3(8) &=-\frac{\p}{10080} \int_0^1 du~   (-1 + u) (20160 - 17016 u - 2584 u^2 + 2386 u^3 + 391 u^4 \NO\\[5pt]
&- 183 u^5 - 15 u^6 + 5 u^7)=\frac{\p}{2} \cdot \frac{8051}{5670}~,\\[10pt]
\label{f4}
\wt F_4(10) &=-\frac{\p}{725760} \int_0^1 du~  (u-1) (-1814400 + 1495728 u + 246348 u^2 - 223072 u^3 \NO\\[5pt]
&- 45587 u^4 +  20395 u^5 + 2839 u^6 - 881 u^7 - 56 u^8 + 14 u^9)=\frac{\p}{2} \cdot \frac{1339661}{748440}~.
\end{align}
The $a$-anomaly coefficients obtained from \eqref{f1}-\eqref{f4} are in complete agreement with the corresponding values in \cite{Cappelli:2000fe,Giombi:2016pvg} which were obtained through the zeta function regularization methods. 

For  $d\ne 2p+2$, but $d$ even, the theory is not conformal. Therefore, in these cases $(-1)^{d/2}(2/\p) \wt F$ cannot be interpreted as an anomaly coefficient. However, it still fixes the coefficient of the curvature counterterms in the calculation of the renormalized free energy in dimensional regularization \cite{Giombi:2015haa}. Some values of $\wt F$ in these cases are listed below\footnote{Eq. \eqref{p0} are the values of the non-conformal scalar with the zero-mode removed. Hence, they should not be confused with an anomaly coefficient.}
\begin{align}\label{p0}
\wt F_0(4)& =-\frac{\p}{2} \cdot \frac{29}{90}~,~~\wt F_0(6)=\frac{\p}{2} \cdot \frac{1139}{3780}~,~~\wt F_0(8)=-\frac{\p}{2} \cdot \frac{32377}{113400}~,\\[10pt]
\label{p1}
\wt F_1(6)& =-\frac{\p}{2} \cdot \frac{1271}{1890}~,~~\wt F_1(8)=\frac{\p}{2} \cdot \frac{4021}{6300}~,~~\wt F_1(10)=-\frac{\p}{2} \cdot \frac{456569}{748440}~,\\[10pt]
\label{p2}
\wt F_2(4)& =-\frac{\p}{2} \cdot \frac{209}{90}~,~~\wt F_2(8)=-\frac{\p}{2} \cdot \frac{2603}{2520}~,~~\wt F_2(10)=\frac{\p}{2} \cdot \frac{13228}{13365}~,\\[10pt]
\label{p3}
\wt F_3(4)& =\frac{\p}{2} \cdot 4~,~~\wt F_3(6) =-\frac{\p}{2} \cdot \frac{5051}{1890}~,~~\wt F_3(10)=-\frac{\p}{2} \cdot \frac{5233531}{3742200}~,\\[10pt]\label{p4}
\wt F_4(4)& =-\frac{\p}{2} \cdot 6~,~~\wt F_4(6) =\frac{\p}{2} \cdot \frac{16259}{3780}~,~~\wt F_4(8)=-\frac{\p}{2} \cdot \frac{7643}{2520}~.
\end{align}

For odd $d$, straightforward evaluation of \eqref{freeen2} yields the following results expressed in terms of $\wt F$
 \begin{align}\label{dim3}
\wt F_1(3)=\wt F_0(3)&=-\frac12 \log(2\p \m_0R)+\frac{\z(3)}{4\p^2}~,\\[10pt]\label{dim5}
\wt F_0(5)=\wt F_3(5)&=\frac32 \log (\m_0R)+\frac{1}{2} \log \left(2 \pi ^2\right)-\frac{23 \zeta (3)}{48 \pi ^2}+\frac{\zeta (5)}{16 \pi ^4}~,\\[10pt]\label{12-form}
\wt F_1(5)=\wt F_2(5)&=-\frac{1}{2} \log (2 \pi \m_0R)+\frac{5 \zeta (3)}{16 \pi ^2}+\frac{3 \zeta (5)}{16 \pi ^4}~,\\[10pt]\label{dim7}
\wt F_1(7)=\wt F_4(7)&=\frac32 \log (\m_0R)+\frac12 \log (2\p^2)+\frac{5 \zeta (7)}{64 \pi ^6}-\frac{179 \zeta (3)}{288 \pi ^2}-\frac{17 \zeta (5)}{48 \pi ^4}~,\\[10pt]\label{dim77}
\wt F_2(7)=\wt F_3(7)&=-\frac{1}{2} \log (2 \pi \m_0R)+\frac{49 \zeta (3)}{144 \pi ^2}+\frac{7 \zeta (5)}{24 \pi ^4}+\frac{5 \zeta (7)}{32 \pi ^6}~,\\[10pt]\label{dim9}
\wt F_3(9)=\wt F_4(9)&=-\frac{1}{2} \log (2 \pi \m_0R)+\frac{205 \zeta (3)}{576 \pi ^2}+\frac{91 \zeta (5)}{256 \pi ^4}+\frac{75 \zeta (7)}{256 \pi ^6}+\frac{35 \zeta (9)}{256 \pi ^8}~.
 \end{align}
Trading off the renormalization scale $\m_0$ with the coupling $e$ (and factors of $2\pi$) we find that the expression for the 1-form in \eqref{12-form} matches with the result in \cite{Giombi:2015haa}. Furthermore we see that it is possible to reproduce this result from a 2-form field on $S^5$ as expected from classical Hodge duality.


\section{Discussion: Classical vs. Quantum equivalence}
In $d$ dimensions a $p$-form theory is Hodge dual to a $(d-p-2)$-form theory. For example, in three dimensions a massless vector (which has just one propagating local degree of freedom) can be dualized (at least locally) to a scalar via the relation 
\be 
F_{\m\n}\sim\e_{\m\n\r}\pa_\r \f~.
\ee
This is a classical equivalence. At least for free theories, this equivalence is expected to hold also at the full quantum level. However, looking at the formal expression of the partition function in Eq. \eqref{partitionfunc}, it is not obvious that there exists a quantum equivalence (i.e., equivalence of partition functions $Z_p=Z_{d-p-2}$) between these theories. The partition function depends in a non-trivial manner on the spectrum of the various Laplacians defined on the sphere\footnote{Ref. \cite{Beasley:2014ila} carries out a detailed study of the duality between a scalar and a vector on a generic Riemannian three-manifold having non-tivial cohomology.}. 

\begin{figure}[ht]
\begin{center}
\includegraphics[height=0.28\textheight]{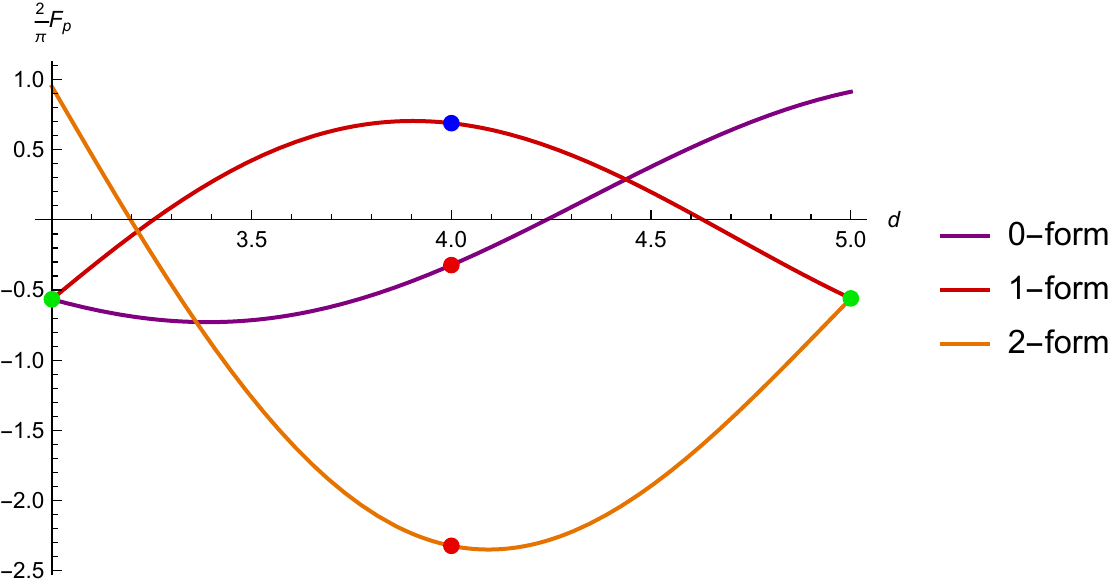}
\caption{Plot of the smoothly interpolating function $(2/\p)\wt F_p(d)$ for $p=0,1,2$ (the dimensionless quantity $\m_0R$ has been set to 1). The green points at $d=3 \tx{ and }5$ indicate the equivalences of the dual theories at the quantum level. The red points at coordinates $(4, -\frac{29}{90})$ and $(4,-\frac{209}{90})$ instead indicate a mismatch between a 0-form and a 2-form in 4 dimensions. The blue dot at $(4,\frac{31}{45})$ is the $a$-anomaly coefficient of the 1-form.}
\label{fer1}
\end{center}
\end{figure}

In the previous section we have explicitly calculated $\wt F_p(d)$ from the general expression in Eq. \eqref{freeen2}. The results in equations \eqref{dim3}-\eqref{dim9} implies that classical equivalence also holds at the quantum level for various Hodge dual pairs in odd dimensions. On the contrary, for $d$ even the results in equations \eqref{p0}-\eqref{p4} imply (at least superficially) that Hodge duality breaks down at the quantum level. A plot of the function $(2/\p)\wt F(d)$ in Fig. \ref{fer1} summarizes the scenarios in various dimensions for $p=0,1,2$.

Let us discuss the case of even $d$ in more detail. For even $d$, the values of $(-1)^{d/2}(2/\p) \wt F$ for the Hodge dual pairs mildly disagree. For example, consider the pair of (classically) Hodge dual theories $(p,d-p-2)=(0,2), (1,3),(2,4)$ in dimensions $4,6,8$ respectively. From Eqs. \eqref{p0}-\eqref{p4} we see that the difference in values of $(-1)^{d/2}(2/\p) \wt F$ is $-2,2,-2$ respectively. The $(0,4)$ Hodge dual pair in $6$ dimensions has a difference of $-4$, and so on. 

The quantum inequivalence between a 2-form and a scalar in 4 dimensions has been discussed in \cite{Bern:2016,Bern:2015xsa,Duff:1980qv,Larsen:2015aia} (and references therein). In particular, our result for $\wt F_0(4),\wt F_2(4)$ and $\wt F_3(4)$ are in agreement with discussions in \cite{Bern:2016,Duff:1980qv}. We recall that for non-conformal $p$-form theories in even $d$ , $(-1)^{d/2}(2/\p) \wt F$ yields the $1/\e$ pole in dimensional regularization and fixes the curvature counterterm coefficient in the calculation of the renormalized free energy. In \cite{Duff:1980qv} it was remarked that the difference between counterterm for a 2-form and a 0-form in 4 dimension is equal to the Euler characteristic $\chi$. Since for a sphere the Euler characteristic $\chi=2$, we find that our results are consistent with the findings of \cite{Duff:1980qv}. Curiously, we further observe that a 3-form in 4 dimensions (which has no propagating degrees of freedom) has a non-vanishing value of $(-1)^{d/2}(2/\p) \wt F=4$. This is $-2$ times the difference between the $(-1)^{d/2}(2/\p) \wt F$ values of a 2-form and a scalar in 4 dimensions. This is again in agreement with the findings of \cite{Duff:1980qv}. Ref. \cite{Bastianelli:2005vk} gives a general formula for the difference in the value of $(-1)^{d/2}(2/\p) \wt F$ for various values of $p$ and $d$. Our results in various even dimension is in agreement with their general formula in Eq. (5.6).


\section*{Acknowledgements}
I sincerely thank Matteo Bertolini for encouraging me to write up this note. I am grateful to Matteo Bertolini, Giulio Bonelli, Loriano Bonora and Lorenzo di Pietro for their feedbacks on a preliminary draft version. I particularly thank Giulio Bonelli and Lorenzo Di Pietro for critical comments and helpful discussions. I am also grateful to Vladimir Bashmakov, Aditya Bawane, Andrea Cappelli and Grigory Tarnopolsky for useful discussions and/or email correspondence.

\appendix
\section{$q_{p}(u,d)$ for $p=1,2,3,4$}\label{app1}
In this appendix we present the expression of the polynomials $q_{p}(u,d)$ appearing in eq. \eqref{pformcont1} for $p=1,2,3,4$. 
\begin{align}
q_{1}(u,d)=& ~1 - 3 d + d^2 + (4 - 3 d) u + 2 u^2~,\\
q_{2}(u,d)=&~4 - 42 d + 35 d^2 - 
 10 d^3 + d^4 + (60 - 77 d + 16 d^2 + 4 d^3 - d^4) u \NO\\
 &+ (62 - 20 d - 
    14 d^2 + 4 d^3) u^2 + (20 + 7 d - 5 d^2) u^3 + (2 + 2 d) u^4 ~,\\
q_{3}(u,d)= &~-144 - 2520 d + 3104 d^2 - 1470 d^3 + 350 d^4 - 42 d^5 + 2 d^6 \NO\\
 &+ (3840 - 6612 d + 2721 d^2 + 31 d^3 - 228 d^4 + 47 d^5 - 3 d^6) u \NO\\
    &+ (5488 - 3315 d - 969 d^2 + 876 d^3 - 143 d^4 + d^6) u^2 \NO\\
    &+ (2844 + 212 d - 1077 d^2 + 192 d^3 + 24 d^4 - 5 d^5) u^3 \NO\\
    &+ (700 + 447 d - 196 d^2 - 30 d^3 + 9 d^4) u^4 \NO\\
    &+ (84 + 94 d + 3 d^2 - 7 d^3) u^5 \NO\\
    &+ (4 + 6 d + 2 d^2) u^6~,\\
q_{4}(u,d)=& -79488 - 399456 d + 646536 d^2 - 399096 d^3 + 134694 d^4 - 
 27216 d^5  \NO\\
 &+ 3276 d^6 - 216 d^7 + 6 d^8 \NO\\
 &+ (651456 - 1376952 d + 792160 d^2 - 107064 d^3 - 50959 d^4 + 
    23376 d^5  \NO
                    \end{align}
    \begin{align}
    &- 4054 d^6 + 336 d^7 - 11 d^8) u \NO\\
    &+ (1157136 - 
    956560 d - 63860 d^2 + 246060 d^3 - 79268 d^4 + 8422 d^5  \NO\\
    &+  274 d^6 - 114 d^7 + 6 d^8) u^2 \NO\\
    &+ (765936 - 100486 d - 
    304907 d^2 + 113656 d^3 - 2988 d^4 - 3858 d^5  \NO\\
    &+ 540 d^6 - 12 d^7 - d^8) u^3 \NO\\
     &+ (267084 + 111150 d - 108772 d^2 + 2234 d^3 + 
    7142 d^4 - 950 d^5 - 22 d^6 + 6 d^7) u^4 \NO\\
    &+ (54432 + 47420 d - 
    11162 d^2 - 5462 d^3 + 1072 d^4 + 66 d^5 - 14 d^6) u^5 \NO\\
    &+ (6552 + 
    8172 d + 876 d^2 - 796 d^3 - 36 d^4 + 16 d^5) u^6 \NO\\
    &+ (432 +  674 d + 237 d^2 - 14 d^3 - 9 d^4) u^7 \NO\\
    &+ (12 + 22 d + 12 d^2 + 2 d^3) u^8~.
\end{align}


\end{document}

%% file: macros.tex
\def\){\right)}
\def\({\left( }
\def\]{\right] }
\def\[{\left[ }

\def\NO{\nonumber}

\newcommand{\be}{\begin{equation}}
\newcommand{\ee}{\end{equation}}

\def\bea{\begin{eqnarray}}
\def\eea{\end{eqnarray}}

\def\bal#1\eal{\begin{align}#1\end{align}}

\def\bald{\begin{aligned}}
\def\eald{\end{aligned}}

\def\beqx{\begin{displaymath}}
\def\eeqx{\end{displaymath}}

\newcommand{\bmat}{\left(\begin{array}}
\newcommand{\emat}{\end{array}\right)}

\def\e{\epsilon}
\def\f{\phi}

\def\l{\lambda}
\def\m{\mu}
\def\n{\nu}

\def\p{\pi}

\def\r{\rho}

\def\z{\zeta}
\def\D{\Delta}

\def\G{\Gamma}

\def\ba{\bbalpha}

\def\cm{{\cal M}}

\def\bo{{\raise-.3ex\hbox{\large$\Box$}}}               
\def\pa{\partial}                                       
\def\face{{\raise.2ex\hbox{$\displaystyle \bigodot$}\mskip-2.2mu \llap {$\ddot
        \smile$}}}                                   
\def\>{\rangle}                                      
\def\<{\langle}                                      

\def\wt#1{\widetilde{#1}}                            
\def\leftrightarrowfill{$\mathsurround=0pt \mathord\leftarrow \mkern-6mu
        \cleaders\hbox{$\mkern-2mu \mathord- \mkern-2mu$}\hfill
        \mkern-6mu \mathord\rightarrow$}        
\def\dvec#1{\vbox{\ialign{##\crcr
        \leftrightarrowfill\crcr\noalign{\kern-1pt\nointerlineskip}
        $\hfil\displaystyle{#1}\hfil$\crcr}}}           

\def\-{\hphantom{-}}

